\documentstyle[12pt,epsf]{article}

\newcommand{\gsim}{\lower.7ex\hbox{$\;\stackrel{\textstyle>}{\sim}\;$}}
\newcommand{\lsim}{\lower.7ex\hbox{$\;\stackrel{\textstyle<}{\sim}\;$}}

\def\prc#1#2#3{\mbox{Phys. Rep. {\bf #1} (#2) #3}}

\def\ijmp#1#2#3{\mbox{Int. J. Mod. Phys. {\bf A#1} (#2) #3}}

\def\beq{\begin{equation}}
\def\eeq{\end{equation}}
\def\bea{\begin{eqnarray}}
\def\eea{\end{eqnarray}}
\def\bq{\begin{quote}}
\def\eq{\end{quote}}

\def\bq{\begin{quote}}
\def\eq{\end{quote}}

\begin{document}

\baselineskip 24pt
\newcommand{\sheptitle}
{Fine-Tuning Constraints on Supergravity Models}

\newcommand{\shepauthor}
{M. Bastero-Gil$^1$, G. L. Kane$^2$ and S. F. King$^1$ }

\newcommand{\shepaddress}
{$^1$Department of Physics and Astronomy,
University of Southampton, Southampton, SO17 1BJ, U.K.\\
$^2$Randall Physics Laboratory, University of Michigan,
Ann Arbor, MI 48109-1120}

\newcommand{\shepabstract} {We discuss fine-tuning constraints
on supergravity models. The tightest constraints come from the
experimental mass limits on two key particles: the lightest CP
even Higgs boson and the gluino. We also include
the lightest chargino which is relevant when universal gaugino masses
are assumed.
For each of these particles we show how fine-tuning increases 
with the experimental mass limit, for four types of supergravity
model: minimal supergravity, no-scale supergravity (relaxing the 
universal gaugino mass assumption), D-brane models
and anomaly mediated supersymmetry
breaking models. Among these models, the D-brane model is less
fine tuned.The experimental propects for an early discovery  
of Higgs and supersymmetry at LEP and the Tevatron are discussed
in this framework.}

\begin{titlepage}
\begin{flushright}
hep-ph/9910506\\
SHEP 99/18
\end{flushright}

\begin{center}
{\large{\bf \sheptitle}}
\bigskip \\ \shepauthor \\ \mbox{} \\ {\it \shepaddress} \\ \vspace{.5in}
{\bf Abstract} \bigskip \end{center} \setcounter{page}{0}
\shepabstract
\begin{flushleft}
\today
\end{flushleft}
\end{titlepage}

When should physicists give up on low energy supersymmetry?
The question revolves around the issue of how much fine-tuning one
is prepared to tolerate. Although fine-tuning is not a well defined concept,
the general notion of fine-tuning is unavoidable 
since it is the existence of fine-tuning in the
standard model which provides the strongest motivation for low energy 
supersymmetry, and the widespread belief that superpartners should
be found before or at the LHC. Although a precise measure of 
{\em absolute } fine-tuning is impossible,
the idea of {\em relative fine-tuning}
can be helpful in selecting certain models and regions
of parameter space over others. 

For example, in a recent paper
two of us investigated non-universal soft parameter space and
concluded that (i) lowering the high energy gluino mass $M_3$ relative
to $M_{1,2}$ reduced fine-tuning (because fine-tuning is mostly
sensitive to $M_3$), (ii) having certain relations between the
soft parameters at the high energy scale such as one between the
up-type Higgs doublet mass and the gluino mass $m_{H_U}\approx 2M_3$,
can reduce fine-tuning\footnote{Another example of how to reduce
fine-tuning not mentioned in
\cite{KK} is to increase $M_2$ for fixed $M_3$, due to the
cancellation effect. } \cite{KK}. 
These results follow from our
observation that fine-tuning is mainly dominated by $M_3$,
and this dominant contribution can be partly cancelled by
negative contributions from other soft parameters, as can
be clearly seen from the expansion of the $Z$ mass
in terms of high energy input parameters \cite{KK},
for example for $\tan \beta =2.5$ we find
\bea
\frac{M_Z^2}{2} =  
& & \mbox{} - .87\,\mu^2(0) +
3.6\,{M_3^2(0)}- .12\,{ M_2^2(0)} + .007\,{M_1^2(0)}  \nonumber \\
& & \mbox{} - .71\, {m_{H_U}^2(0)} + .19\,{ m_{H_D}^2(0)}
+ .48\,({ m_Q^2(0)} + \,{m_U^2(0)}) 
\nonumber \\
 & & \mbox{}  - .34\,{A_t(0)}\,{M_3(0)} - .07\,{ A_t(0)}\,{M_2(0)} 
- .01\,{A_t(0)}\,{M_1(0)}
+ .09\,{ A_t^2(0)}\nonumber \\
 & & \mbox{} + .25\,{M_2(0)}\,{M_3(0)}+ .03\,{M_1(0)}\,{M_3(0)} 
+ .007\,{M_1(0)}\,{M_2(0)} \,
\label{MZ}
\eea
where we have implicitly assumed  all the soft breaking parameters to
be real, neglecting the phases\footnote{ In the most general situation
with complex soft breaking terms, the phases will enter in the
subleading cross-terms in Eq. (\ref{MZ}).}.  
One implication of the fact that fine-tuning is dominated
by $M_3$ is the fact that the soft scalar
masses can be larger than $M_3$ without increasing fine-tuning,
a fact which has recently been emphasised in 
the framework of minimal supergravity in ref.\cite{recent}.

In this paper we shall extend the discussion in ref.\cite{KK} in two
ways. Firstly we shall study fine-tuning in various supergravity
models: minimal supergravity, no-scale supergravity (relaxing the 
universal gaugino mass assumption), D-brane models
and anomaly mediated supersymmetry
breaking models (AMSB). The common feature of this class of models is that
they involve a large mass scale of order the unification scale
say $M_U\sim 2 \times 10^{16}$ GeV, and supersymmetry breaking is
mediated via some sort of hidden sector supergravity mechanism.
Thus our analysis does not extend to either gauge-mediated
supersymmetry breaking models, or models where the string scale is
lowered beneath the unification scale, although it may be lowered to
the unification scale. The reason why we choose these models is that
they contain the largest mass hierarchy, and hence face the most severe
fine-tuning constraints in general. These
models also preserve the gauge unification success
most simply and directly.

Secondly we focus on the key particles whose
experimental mass limits lead most sensitively to increases
in fine-tuning. Clearly fine-tuning is not sensitive to
squark and slepton masses which can be increased
substantially due to the insensitivity of the $Z$ mass formula
in Eq.\ref{MZ} to soft scalar masses. By contrast the lightest 
CP even Higgs mass is a very sensitive probe of fine-tuning, 
as we emphasised previously \cite{KK}, and it is obvious from
the foregoing discussion that the gluino mass itself is also
a sensitive probe. Although we showed \cite{KK} that the chargino mass
is only a sensitive probe of fine-tuning 
if one assumes universal
gaugino masses, we shall nevertheless include it for illustrative
purposes.

The implicit sensitivity of
the Z mass coming from changes in $\tan \beta$ as a result of
small variations in the high energy inputs, does not appear in Eq.\ref{MZ}.
This is addressed by the master formula of Dimopoulos and Giudice
\cite{giudice} which yields a fine-tuning parameter which corresponds
to the fractional change in the Z mass squared per unit fractional
change in the input parameter,
\beq
\Delta_a=abs \left( \frac{a}{M_Z^2}\frac{\partial M_Z^2}{\partial a}\right)
\eeq
for each input parameter $a$ in the model of interest.
The fine-tuning is then simply the maximum value of $\Delta_a$ over
all the input parameters.
Although there are many more sophisticated measures of fine-tuning
available \cite{giudice,earlier,price,Wright}, this basic measure of
fine-tuning is adequate for our purposes of comparing relative fine-tunings
amongst different models.

The models we consider, and the corresponding input parameters given
at the unification scale,  are listed below: 
\begin{enumerate}

\item Minimal supergravity \cite{recent}.
\bea
a_{msugra}\in \{m_0^2, M_{1/2}, A(0), B(0), \mu(0)\}\,,
\label{msugra}
\eea 
where as usual $m_0$, $M_{1/2}$ and $A(0)$ are the universal scalar mass,
gaugino mass and trilinear coupling respectively, $B(0)$ is the
soft breaking bilinear coupling in the Higgs potential and $\mu(0)$ is
the Higgsino mass parameter. 

\item No-scale supergravity \cite{noscale} with non-universal gaugino
masses\footnote{This is in fact a new model not
previously considered in the literature, 
although the no-scale model with universal gaugino masses
is of course well known. As in the usual no-scale model,
this model has the attractive feature that 
flavour-changing neutral currents at low energies are very suppressed,
since all the scalar masses are generated by radiative corrections,
via the renormalisation group equations, which only depend on the
gauge couplings which are of course flavour-independent.}
\bea
a_{no-scale}\in \{M_1(0), M_2(0), M_3(0), B(0), \mu(0)\}
\label{noscalesugra}
\eea 

\item D-brane model \cite{dbrane}.
\bea
a_{D-brane}\in \{m_{3/2}, \theta , \Theta_1, \Theta_2, \Theta_3,
 B(0), \mu(0)\}\,,
\label{Dbrane}
\eea 
where $\theta$ and $\Theta_i$ are the goldstino angles, with
$\Theta_1^2+\Theta_2^2+\Theta_3^2=1$, and $m_{3/2}$
is the gravitino mass. The gaugino masses are given by
\bea
M_1(0)  =  M_3(0) & = & \sqrt{3}m_{3/2}\cos \theta \Theta_1 e^{-i\alpha_1}
\nonumber \,,\\
M_2(0) & = & \sqrt{3} m_{3/2}\cos \theta \Theta_2 e^{-i\alpha_2} \,,
\eea
and there are two types of soft scalar masses
\bea
m_{5152}^2 & = & m_{3/2}^2
[1-\frac{3}{2}(\sin^2 \theta +\cos^2 \theta \Theta_3^2) ]
\nonumber \,,\\
m_{51}^2 & = & m_{3/2}^2[1-3\sin^2 \theta] \,,
\eea

\item Anomaly mediated supersymmetry breaking \cite{amsb}.
\bea
a_{AMSB}\in \{m_{3/2}, m_0^2, B(0), \mu(0)\}
\label{AMSB}
\eea 

\end{enumerate}

Our numerical results are based on two-loop renormalisation group
running of gauge\footnote{When running the gauge
couplings we have included complete threshold effects at order 1-loop
\cite{thresholds} 
and used the step-function approximation in the 2-loops
coefficients.}, third generation Yukawa couplings and soft mass
parameters \cite{2soft}.   
The initial values of the Yukawa couplings are determined by the
values of the third generation fermion masses\footnote{We have
included one-loop susy threshold corrections, QCD and electroweak
corrections  when converting pole mass
values to running mass values at the $m_Z$ scale \cite{pierce}.}. The
input soft mass parameters are then 
chosen in order to get electroweak symmetry breaking and the $m_Z$
scale given by the minimisation conditions of the one-loop corrected
Higgs potential \cite{effpot1,effpot2}, 
\begin{equation}
\frac{m_Z^2}{2}=\frac{m_{H_D}^2 - m_{H_U}^2 \tan \beta^2 -\Delta_Z^2}
{\tan \beta^2-1} - \mu^2\,, \label{min} 
\end{equation}
where $\tan\beta= \langle H_U \rangle / \langle H_D
\rangle$, $\Delta_Z^2$ is the one-loop contribution, and the
parameters in Eq. (\ref{min}) are evaluated at $m_Z$.  
In practice, for the numerical calculations we use as input
$\tan\beta$ and sign$(\mu)$ (we always take $\mu > 0$) and obtain
$\mu(0)$ and $B(0)$ from the minimisation conditions. 

Our main results are shown in Figs.1-4, corresponding to SUGRA
models 1-4 above. 
The results are shown for three values of $\tan \beta = 2,3,10$, corresponding
to three sets of curves from top left to bottom right, respectively.
In each case we plot the maximum sensitivity
parameter $\Delta^{max}$ as a function of particle mass, for
the lightest CP even Higgs boson (short dashes), the lighter
chargino (long dashes) and the gluino (solid).
The lightest CP even Higgs boson is calculated using the one-loop
RG-improved effective potential approach \cite{higgsRG}, which
includes the leading two-loop corrections to the Higgs
mass\footnote{The expected accuracy in the computed Higgs mass is
estimated to be $\sim 2$ GeV. A different approach to the calculation of
the Higgs mass can be founded in Ref. \cite{hollik}.}. The gluino mass
also includes the corrections due to gluon/gluino and quark/squark loops
 \cite{gluinorc,pierce}. 

\begin{figure}[ht]
\epsfxsize=15cm
\epsfysize=8cm
\hfil \epsfbox{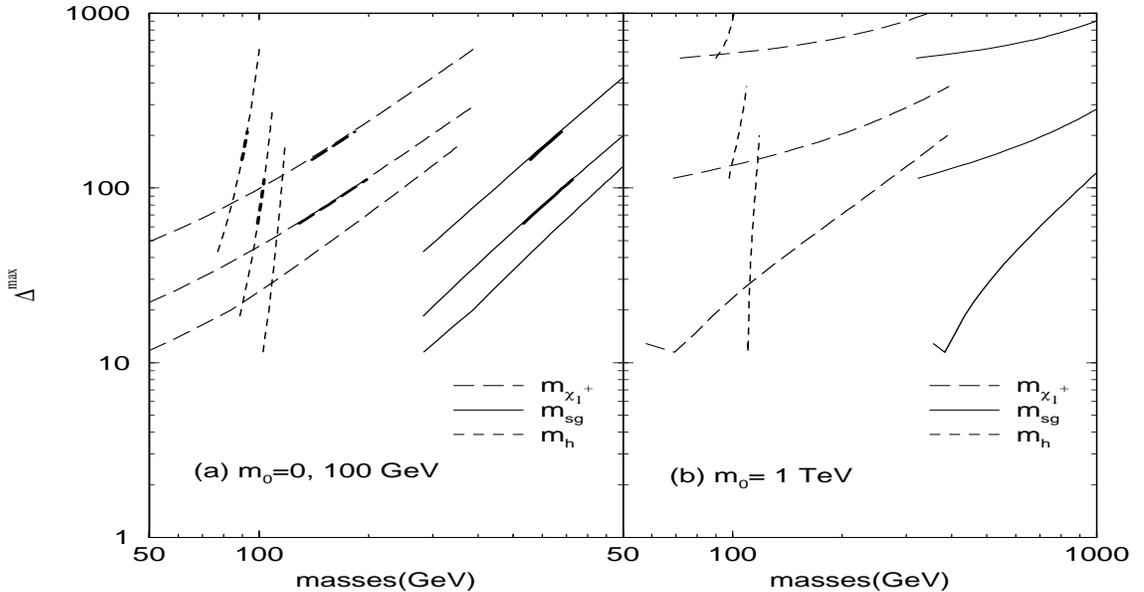} \hfil
\caption{ \small{Results for the minimal SUGRA model.
The maximum sensitivity parameter $\Delta^{max}$ is plotted 
as a function of the lightest CP even Higgs mass (short dashes),
gluino mass (solid line) and lightest chargino (long dashes).
For each particle type, the three sets of curves correspond to 
$\tan\beta$=2, 3, 10, from top left to bottom right, respectively. 
In panel (a) the shorter, thicker lines
correspond to $m_0=0$, while the longer lines are those for $m_0=100$
GeV. In panel (b) the results correspond to $m_0=1000$ GeV.}}
\end{figure}

In Fig.1 we give the fine-tuning results for mSUGRA.
The present LEP2 mass limits of around 100 GeV
on the Higgs and chargino compete
for providing the tightest fine-tuning constraint,
while the current Tevatron gluino mass limit of around 250 GeV 
provides a slightly less severe limit. For a Higgs mass of 100 GeV,
$\tan \beta =10$ allows fine-tuning to stay at around 10,
but as the Higgs mass increases it rapidly overtakes the
chargino mass in importance and as it approaches 110 GeV fine-tuning rises
steeply to 100. Comparing Fig.1(a) with $m_0=100$ GeV, to 
Fig.1(b) with $m_0=1000$ GeV we see that for $\tan \beta =10$ 
the curves are very similar, as emphasised
in ref.\cite{recent}. However we emphasise that for lower values of
$\tan \beta $ fine-tuning increases substantially as $m_0$
is increased from 100 GeV to 1000 GeV in mSUGRA.
Also in Fig.1(a) we show results for no-scale mSUGRA
with $m_0=0$ for $\tan \beta =2,3$ seen as short lines
almost superimposed over the $m_0=100$ GeV lines.
The reason why the no-scale lines are so short is that if $M_{1/2}$
is too small the right-handed slepton falls below its experimental
limit of 88 GeV, while if $M_{1/2}$ is too large it becomes the LSP.
Thus there is only a narrow allowed window for $M_{1/2}$ which in the
case of $\tan \beta =10$ is non-existent. 

In Fig.2 we give results for a generalised version of no-scale mSUGRA 
hitherto not considered in the literature in which 
$m_0=0$ as usual, but now we allow the gaugino masses to 
be non-universal. For definiteness we take $M_1(0)=M_2(0)$,
but allow these gaugino masses to be different from the
high energy gluino mass $M_3(0)$. The first point to make is that
by relaxing gaugino mass universality, a larger parameter space is
opened up and the constraints which forced $M_{1/2}$ into a small
allowed range in the no-scale mSUGRA model are now replaced by large
allowed regions in non-universal gaugino mass space. For example
taking $M_2(0)=M_1(0)=250$ GeV
in Fig.2(a) we see a large range of $M_3(0)$ is allowed.
Also we find that fine-tuning is generally smaller in this model
than mSUGRA for $\tan \beta = 10$.
The reason is that although the Higgs curves in Fig.2(a) are
very similar to those in Fig.1(a), the chargino curves are very
different. In the no-scale model with non-universal gaugino masses
a chargino mass limit
of around 100 GeV implies a fine-tuning of between 10 and 20,
almost independently of $\tan \beta$, whereas in the conventional
no-scale model the corresponding fine-tuning is between 20 and 100.

\begin{figure}[ht]
\epsfxsize=15cm
\epsfysize=8cm
\hfil \epsfbox{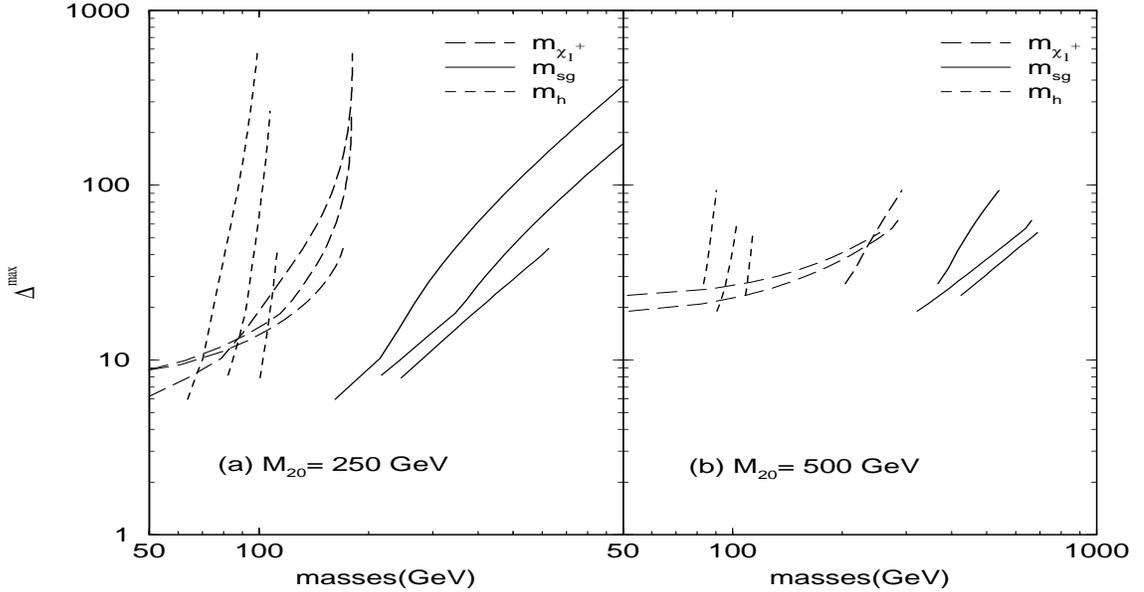} \hfil
\caption{\small{Results for the 
no-scale with non-universal gaugino masses. 
The maximum sensitivity parameter $\Delta^{max}$ is plotted 
as a function of the lightest CP even Higgs mass (short dashes),
gluino mass (solid line) and lightest chargino (long dashes).
For each particle type, the three sets of curves correspond to 
$\tan\beta$=2, 3, 10, from top left to bottom right, respectively. 
In panel (a) we fix $M_2(0)=250$ GeV, while in panel (b) 
$M_2(0)=500$ GeV.}} 
\end{figure}

The kinks in the gluino curves in Fig.2(a) correspond to
$\Delta_{M_3(0)}$ being replaced by $\Delta_{\mu(0)}$  as
the largest fine-tuning parameter as $M_3(0)$ (and thus $\mu(0)$) is
increased. For  $M_3(0) < M_2(0)$ a
partial cancellation occurs in Eq. (\ref{MZ}) between $M_3(0)$ and
$M_2(0)$ \cite{KK}, which renders the fine-tuning for $\mu(0)$ small.  
Because of that, in the region where the chargino is lighter and
mainly higgsino fine-tuning is quite insensitive to its mass. 
This can be seen clearly in Fig.2(b), where  we show results for
$M_2(0)=M_1(0)=500$ GeV. Interestingly for $\tan \beta = 3,10$ the
chargino curves are almost flat, due to its Higgsino nature, while for
$\tan \beta =2 $ the curve is much steeper. 
However, for a chargino mass of order 100 GeV, the overall fine-tuning
is larger than in Fig. 2(a). This corresponds to the increase in
$M_3(0)$ when increasing $M_2(0)$  as required by Eq. (\ref{MZ}). 
For $\tan \beta = 2$ the curves are cut in the region of a light
chargino because the lightest
stop  falls below its experimental lower bound\footnote{The
same effect can be seen for the D-brane model in Fig. (3).} of around 90
GeV. For the three values of $\tan \beta$ considered, an upper limit 
on $M_3(0)$ is set by the requirement that the lightest neutralino
mass does not exceed the slepton mass.

\begin{figure}[ht]
\epsfxsize=15cm
\epsfysize=8cm
\hfil \epsfbox{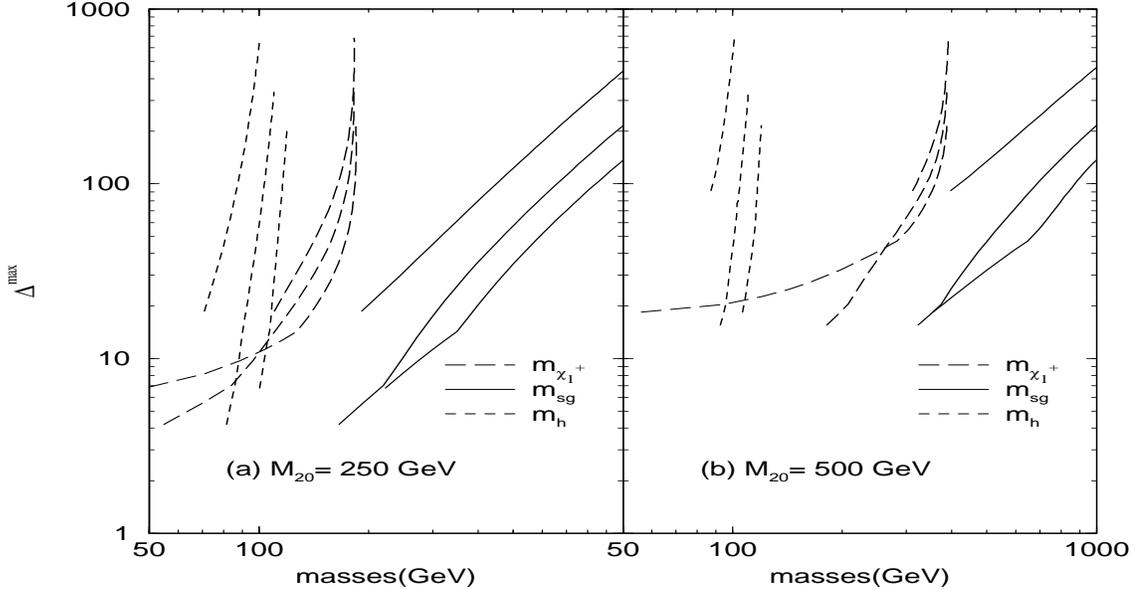} \hfil
\caption{{\small Results for the D-brane model. 
The maximum sensitivity parameter $\Delta^{max}$ is plotted 
as a function of the lightest CP even Higgs mass (short dashes),
gluino mass (solid line) and lightest chargino (long dashes).
For each particle type, the three sets of curves correspond to 
$\tan\beta$=2, 3, 10, from top left to bottom right, respectively.
In panel (a) we fix $M_2(0)=250$ GeV, while in panel (b) 
$M_2(0)=500$ GeV.}}
\end{figure}

In Fig.3 we give results for a D-brane scenario, where
we take the Goldstino angles $\cos \theta =1$ and $\Theta_3 = 0$.
We also set all the scalar masses
equal to the universal value
$m_{5152}^2$ at the high energy scale\footnote{Other choices of the
Goldstino angles or the scalar masses will affect 
mainly the low energy values of the scalar masses, and not so much
those of the gauginos (a change in $\cos\theta$ can be compensated by a
rescaling of the gravitino mass). This may change the region of the
parameter space allowed by the experimental constraints, but it will
leave practically unchanged the conclusions on fine-tuning.
} $M_U$. 
The gaugino masses are again non-universal 
but now $M_1(0)=M_3(0)$ and the ratio of these masses
to $M_2(0)$ is controlled by the Goldstino angles
$\Theta_1$ and $\Theta_2$.
These are constrained to lie along a unit circle, and
thus we have only the freedom to change their ratio
$\Theta=\Theta_1/\Theta_2$ when moving along the
circumference. Therefore, we compute the fine-tuning for $\Theta$
instead of those for $\Theta_1$ and $\Theta_2$. 
The results in Fig.3(a) for $M_2(0)=250$ GeV
are quite similar to those in Fig.2(a), and imply 
a similarly low fine-tuning.
In Fig.3(b) the choice $M_2(0)=250$ GeV now leads to
larger allowed regions than in Fig.2(b) due to the
presence of a non-zero scalar mass, with the charginos
being now significantly heavier due to their gaugino
component. Now the parameters that compete
 to give the largest fine-tuning are $\mu(0)$ and
$\Theta$, and the kink in the gluino curves is due to
$\Delta_{\Theta}$ being replaced by $\Delta_{\mu(0)}$ as the maximum
sensitivity parameter. The other functions $\Delta_{m{3/2}}$ and
$\Delta_\theta$  can become comparable but not dominant. 
As in the generalised no-scale model,  the
maximum fine-tuning will be 
insensitive to a light chargino when this is mainly higgsino. 

\begin{figure}[ht]
\epsfxsize=15cm
\epsfysize=8cm
\hfil \epsfbox{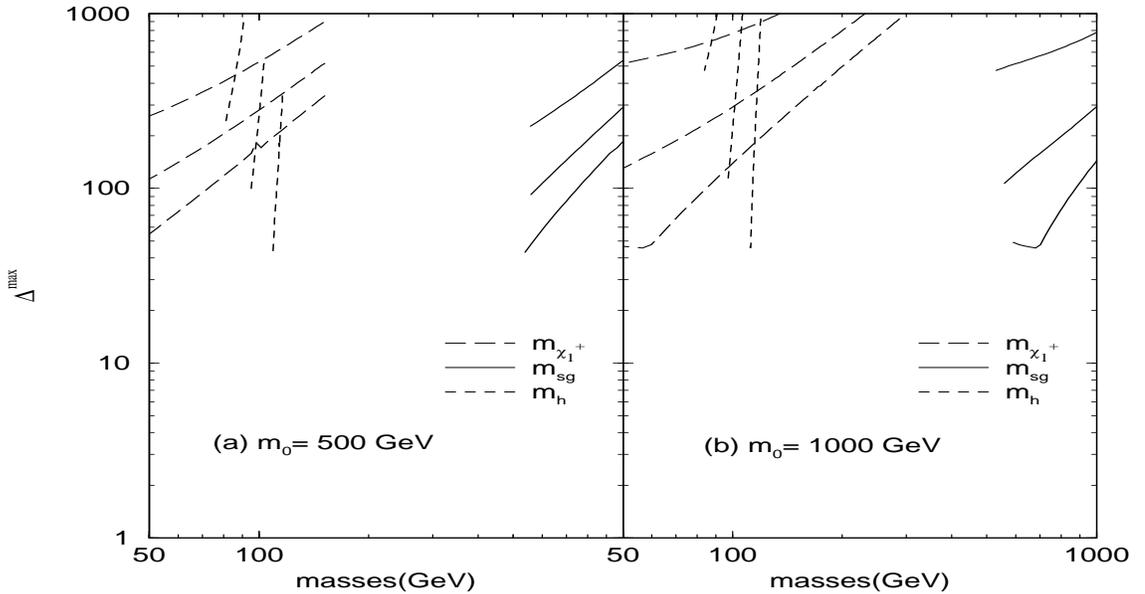} \hfil
\caption{{\small Results for the
anomaly mediated supersymmetry breaking model.
The maximum sensitivity parameter $\Delta^{max}$ is plotted 
as a function of the lightest CP even Higgs mass (short dashes),
gluino mass (solid line) and lightest chargino (long dashes).
For each particle type, the three sets of curves correspond to 
$\tan\beta$=2, 3, 10, from top left to bottom right, respectively.
In panel (a) we fix $m_0=500$ GeV, while in panel (b) 
$m_0=1000$ GeV.}}
\end{figure}

In Fig.4 we give results for the AMSB model.
In the minimal AMSB model where $m_0=0$ the sleptons are
predicted to have negative mass squared, so we have followed
the common proceedure of simply adding a universal scalar mass
squared $m_0^2$ by hand, ensuring that it is large enough to
ensure acceptable slepton masses. In Fig.4(a) we choose
$m_0=500$ GeV, and in Fig.4(b) we take $m_0=1000$ GeV.
In both cases the fine-tuning is dominated by 
$\Delta_{\mu(0)}$, and is much larger than the other models
considered. Typically the value of $\mu(0)$ required by electroweak
symmetry breaking is $O$(1 TeV) or larger in this models.  

Comparing the results for all the models in Figs.1-4 it is seen that 
there is slightly less fine-tuning associated with particle masses
in the D-brane model than in the other models. However it is also
apparent that the results for 
the no-scale model with non-universal gaugino masses are very similar
to the D-brane scenario. The common feature of both these models is
non-universal gaugino masses, and the reasons for the reduced 
fine-tuning are essentially those emphasised in ref.\cite{KK}
(namely that fine-tuning is most sensitive to $M_3(0)$ and so 
$M_3(0)<M_2(0)$ in general reduces fine-tuning.)
However, in the D-brane model there is additionally 
the possibility of cancellations among different input
parameters which help to lower the different fine-tuning parameters. 
For example, using one-loop 
semi-analytic solutions to the renormalisation group equations
\cite{carena} and neglecting one-loop effective potential
contributions, we find the approximate expressions for $\tan\beta=3$:
\bea
\Delta_{m_{3/2}} &\approx& \tilde{m}_{3/2}^2 \left| \cos^2\theta ( 120.67 
\Theta_1^2 -8.15 \Theta_2^2 + 10.13 \Theta_2 \Theta_1) -0.32 (1-3
\cos\theta^2) \right| \label{dm32} \\ 
\Delta_{\mu(0)} &\approx& \left| 5.12 + \tilde{m}_{3/2}^2 \cos^2\theta
(-135.93 \Theta_1^2 +6.85 \Theta_2^2 - 11.40 \Theta_2 \Theta_1 )
\right. \nonumber \\  
 & &\left. + 1.14 \tilde{m}_{3/2}^2 (1-3 \cos^2\theta) \right|
\label{dmu} \,, \\
\Delta_{\Theta} &\approx& \tilde{m}_{3/2}^2 \Theta_1 \Theta_2 \left|
\cos^2\theta ( 128.81 \Theta_1 \Theta_2 + 5.07 (\Theta_2^2 - \Theta_1^2))
 \right| \label{dmx} \,, \\ 
\Delta_{\theta} &\approx& \tilde{m}_{3/2}^2 \cos^2\theta \left|  112.14 
\Theta_1^2 -8.15 \Theta_2^2 + 8.97 \Theta_2 \Theta_1  +0.32  \right|
\label{dma} \,, 
\eea
where $\tilde{m}_{3/2}$ is the gravitino mass scaled by $m_Z$, and we
have kept the dependence on $\cos\theta$ but taken $\Theta_3=0$. From
the above  expressions it is clear that choosing appropiate values for the
goldstino angles, $\Delta_{m_{3/2}}$ might be arbitrarily small even for
very large values of the gravitino mass, and similarly for 
$\Delta_{\mu(0)}$ and $\Delta_{\theta}$.  

\begin{figure}[t]
\epsfxsize=15cm
\epsfysize=15cm
\hfil \epsfbox{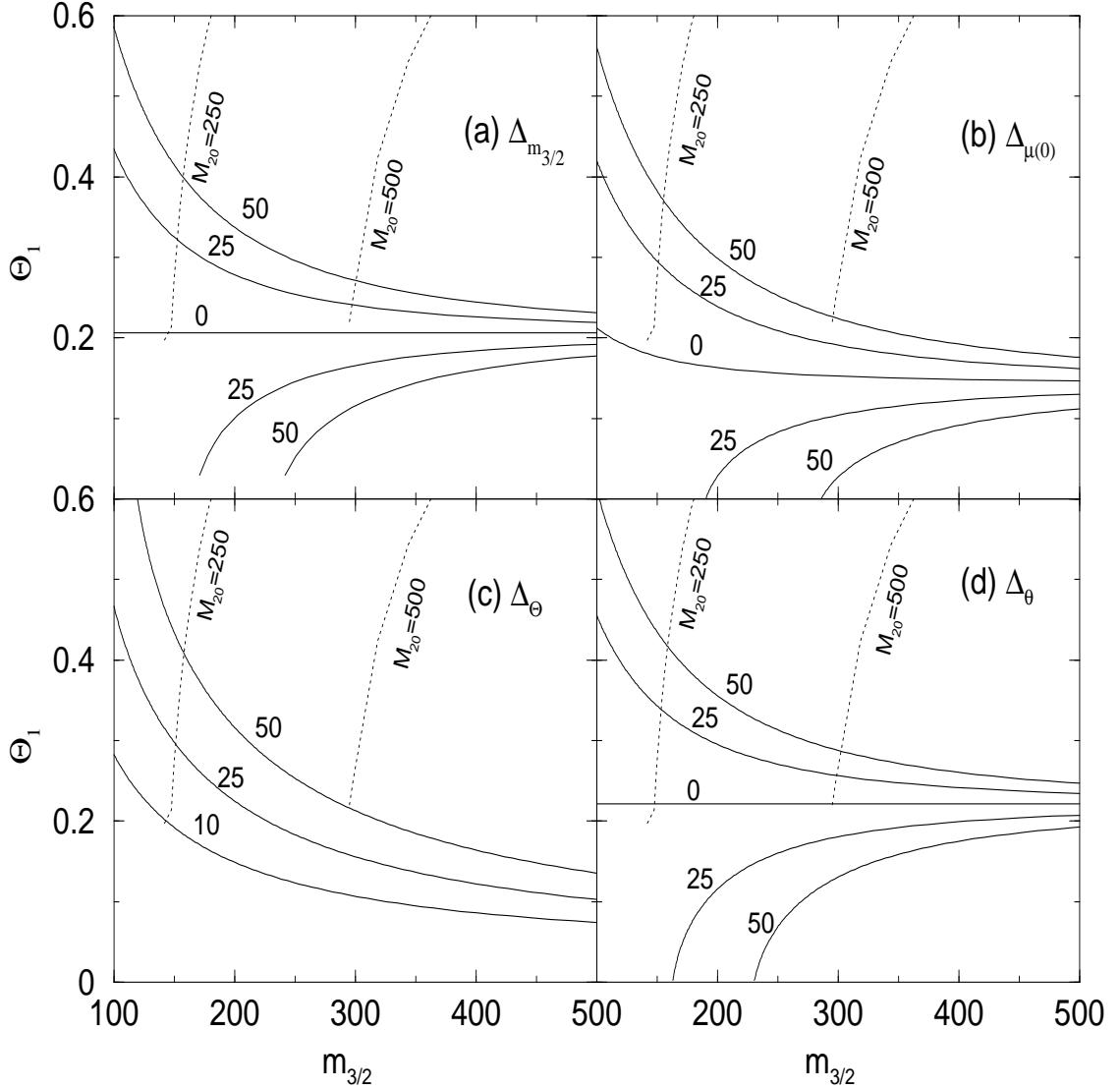} \hfil
\caption{{\small Contours of constant (a) $\Delta_{m_{3/2}}$, 
 (b) $\Delta_{\mu(0)}$, (c) $\Delta_\Theta$, and (d) $\Delta_\theta$,
 given by the approximate one-loop  
expressions Eq. (\ref{dm32} - \ref{dma}), for the D-brane model and
$\tan\beta=3$. We have fixed $\cos\theta=1$ and $\Theta_3=0$. The
 dotted lines are the contours of constant $M_{20}$=250, 500 GeV.}}
\label{demm32}
\end{figure}

In Fig. (\ref{demm32}) we
have plotted the contours of constant $\Delta_i$  given in
Eqs. (\ref{dm32}-\ref{dma}). 
Not suprisingly, all the contours show a hyperbolic behavior: they
would more or less follow the curves of constant $M_3(0)$, with 
fine-tuning  increasing with the gluino mass. 
We have also included the contours
of constant $M_2(0)$ for the values considered in Fig. (3). 
Although $\Delta_{\mu(0)}$,
$\Delta_{m_{3/2}}$ and $\Delta_{\theta}$ are all simultaneously
small for $\Theta_1 \simeq 0.2$, parts of this region are
experimentally excluded due to $\mu $ becoming to small and hence
the lightest chargino becoming too light.
Nevertheless, there are allowed regions in the plane $m_{3/2}-\Theta_i$,
corresponding to a light gluino,  where $\Delta_{\mu(0)}$,
$\Delta_{m_{3/2}}$ and $\Delta_{\theta}$ are all simultaneously
small, and the maximum sensitivity would be given by $\Delta_{\Theta}$. 
We may try to play with the values of either $\cos\theta$ or
$\Theta_3$ in order to find some region where $all$ the fine-tuning
is small. However, reducing (increasing) $\cos\theta$ ($\Theta_3$)
the slepton masses tend to diminish, making it difficult if not impossible
to fulfill the experimental constraints on the SUSY
masses\footnote{For example, for $\cos\theta < 0.5 $ and 
$\tan\beta >3$, the parameter space compatible with experiments shrinks to
nothing.}.

Any conclusions which are drawn from fine-tuning are always 
subject to caveats, disclaimers and health warnings.
A precise value cannot be placed on fine-tuning, since the
definition can always be changed and the question of how much 
fine-tuning is acceptable is subjective.
For this reason we prefer not to give upper bounds on particle
masses based on fine-tuning, but clearly subjective upper bounds
can be read off from our curves, for those inclined to do so.
The main value of our work is to compare different SUGRA models
with each other, and within each SUGRA model to compare different
regions of parameter space, from the point of view of fine-tuning.

In all models, 
fine-tuning is reduced as $\tan \beta$ is increased, with
$\tan \beta =10$ preferred over $\tan \beta =2,3$.
Nevertheless,  
the present LEP2 limit on the Higgs and chargino mass
of about 100 GeV and the gluino mass limit of about 250 GeV
implies that $\Delta^{max}$ is of order 10 or higher. 
The fine-tuning increases most sharply with the Higgs mass.
The Higgs fine-tuning curves are fairly model independent,
and as the Higgs mass limit rises above 100 GeV come to 
quickly dominate the fine-tuning. We conclude that the 
prospects for the discovery of the Higgs boson at LEP2 are good.
For each model there is a correlation between the
Higgs, chargino and gluino mass, for a given value of
fine-tuning. For example if the Higgs is discovered
at a particular mass value, then the corresponding
chargino and gluino mass for each $\tan \beta$ can be 
read off from Figs.1-4. 

The new general features of the results may then be summarised as follows:

\begin{itemize}
\item The gluino mass curves are
less model dependent than the chargino curves,
and this implies that in all models if the fine-tuning
is not too large then the prospects for the 
discovery of the gluino at the Tevatron are good.

\item The fine-tuning due to the chargino mass is model
dependent. For example in the no-scale model 
with non-universal gaugino masses and the
D-brane scenario the charginos may be relatively heavy compared
to mSUGRA.

\item Some models have less fine-tuning than others.
We may order the models on the basis of fine-tuning from
the lowest fine-tuning to the highest fine-tuning:
D-brane scenario $<$ generalised no-scale SUGRA $<$ mSUGRA $<$ AMSB. 

\item The D-brane model is less fine-tuned partly because the gaugino
masses are non-universal, and partly because there are large regions
where $\Delta_{m_{3/2}}$, 
$\Delta_{\mu(0)}$, and $\Delta_\theta$ are all close to zero
(see Fig.\ref{demm32}). However in these regions the fine tuning is 
dominated by $\Delta_\Theta$, and this leads to an inescapable
fine-tuning constraint on the Higgs and gluino mass. 

\end{itemize}

Finally we should comment on the parameter space dependence of
our results. Although the results presented here are for specific
choices of parameters, we have performed a detailed analysis
of the parameter space of these models and found that the 
results are representative of the full parameter space, and
the qualitative conclusions will not change.
We shall present the complete analysis elsewhere \cite{next}.

\begin{center}
{\bf Acknowledgements}
\end{center}

We wish to thank Lisa Everett for useful discussions.
S.K. acknowledges useful discussions at the 1999 Durham Collider
Workshop.

\end{document}